\documentclass[twocolumn,showpacs,preprintnumbers,amsmath,amssymb]{revtex4}
\usepackage{bm}
\usepackage{dcolumn}
\usepackage{graphicx}
\usepackage{color}

\begin{document}
%\draft
\title{Construction and enlargement of dilatonic wormholes by impulsive radiation}
\author{Hiroko Koyama\footnote{Present address: National Astronomical
Observatory of Japan, Mitaka, Tokyo 181-8588, Japan}}
\email{hiroko.koyama@nao.ac.jp}
\author{Sean A. Hayward}
\email{hayward@mm.ewha.ac.kr}
\author{Sung-Won Kim}
\email{sungwon@mm.ewha.ac.kr}
\affiliation{Department of Science Education, Ewha Womans University, Seoul 120-750, Korea}
\date{\today}
\begin{abstract}
The dynamical behavior of traversable wormholes and black holes under impulsive
radiation is studied in an exactly soluble dilaton gravity model. Simple
solutions are presented where a traversable wormhole is constructed from a
black hole, or the throat of a wormhole is stably enlarged or reduced. These
solutions illustrate the basic operating principles needed to construct similar
analytic solutions in full Einstein gravity.
\end{abstract}
 \maketitle

\section{Introduction}
Wormholes are tunnels through space-time, linking otherwise separated regions
of a single universe, or bridges joining two different universes. Since
traversable wormholes were introduced by Morris and Thorne \cite{MT} as
theoretically allowable space-times in Einstein gravity, wormholes have been
pursued as an attractive research topic involving the possibility of rapid
interstellar travel and even time machines, as reviewed by Visser \cite{V}. The
properties of static Morris-Thorne wormholes have been studied by many authors
and it is known even in non-static situations that a negative-energy source is
necessary for traversable wormholes to exist \cite{HV1,HV2,wh,IH}. Assuming
such a source, there are interesting and practical problems of clarifying the
dynamical nature of wormholes, such as how to construct a traversable wormhole
and how to enlarge the throat enough to enable human beings to pass through it.

Not long ago, a unified framework for black holes and traversable wormholes was
proposed \cite{wh}, indicating that they are dynamically interconvertible when
the trapping horizons locally characterizing them \cite{wh,IH,1st,bhd}
bifurcate or merge. This theory was first concretely confirmed in an exactly
soluble model, CGHS two-dimensional dilaton gravity \cite{CGHS} with an
additional negative-energy scalar field to support wormholes \cite{dw}, then in
standard Einstein gravity numerically \cite{SH}. In seeking analytic solutions
in full Einstein gravity, we have found non-static solutions to be difficult to
obtain, except in the idealization of impulsive radiation, where the radiation
is concentrated so as to deliver finite energy and momentum in an instant.

In this paper, we construct some of the simplest models of dynamic wormhole
processes by employing impulsive radiation in a generalized dilaton gravity
model, specifically: wormhole construction from a black hole, wormhole
operation by energy balance and wormhole reduction or enlargement. The last
point addresses a common belief that, while wormholes are possible or even
expected at the Planck scale, large-scale wormholes are unlikely or even
impossible. Actually self-inflating wormholes were recently discovered
numerically \cite{SH}, but to date there have been no concrete examples of
stable wormhole enlargement, where the wormhole size is controlled.

In Sec.~\ref{sec:model} we review the static CGHS \cite{CGHS} black-hole and
HKL \cite{dw} wormhole solutions and generalize the dilaton gravity model. We
find solutions describing the construction of a wormhole from a black hole in
Sec.~\ref{sec:const}. In Secs.~\ref{sec:ope} and \ref{sec:enlarge}, we study
processes to change the throat radius of the wormhole by controlling impulsive
radiation, either from one universe or from both universes. The final section
is devoted to summary.

\section{black holes and wormholes in dilaton gravity}
\label{sec:model} The CGHS two-dimensional dilaton gravity \cite{CGHS} is
generalized by the action \cite{dw}
\begin{equation}\label{action}
 \int_S \mu
\left[ e^{-2\phi } \left( R + 4 (\nabla \phi)^2  + 4\lambda ^2 \right)
 - \frac{1}{2}(\nabla f)^2 + \frac{1}{2}(\nabla g)^2  \right]
\end{equation}
where $S$ is a 2-manifold, $\mu$, $R$ and $\nabla$ are the area form, Ricci
scalar and covariant derivative of a Lorentz 2-metric on $S$, $\lambda$
represents a negative cosmological constant, $\phi$ is a scalar dilaton field,
$f$ is a Klein-Gordon field representing matter and $g$ is a ghost Klein-Gordon
field. The last term is added to the CGHS action in order that $g$ provides the
negative energy densities needed to support a traversable wormhole
\cite{MT,V,wh,HV1,HV2,IH}. By choosing future-pointing null coordinates $(x^+ ,
x^- )$, the line element may be written as
\begin{equation}\label{metric}
  ds^2 = -2 e^{2\rho} dx^+ dx^-.
\end{equation}
Taking the gauge choice $\rho = \phi$ and transforming the dilaton field $\phi$
to $r=2e^{-2\phi}$, the field equations reduce to a simple form: the evolution
equations
\begin{eqnarray}
 \partial_+ \partial_- f &=& 0 \\
 \partial_+ \partial_- g &=& 0 \\
 \partial_+ \partial_- r &=& -4 \lambda^2
\end{eqnarray}
and the constraints
\begin{equation}\label{constraints0}
  \partial_\pm   \partial_\pm  r = ( \partial_\pm  g )^2 - ( \partial_\pm  f )^2.
\end{equation}
The field $r$ plays a similar role to the areal radius in spherical symmetry
\cite{cc,1st}.

\begin{figure}
\includegraphics[height=28mm]{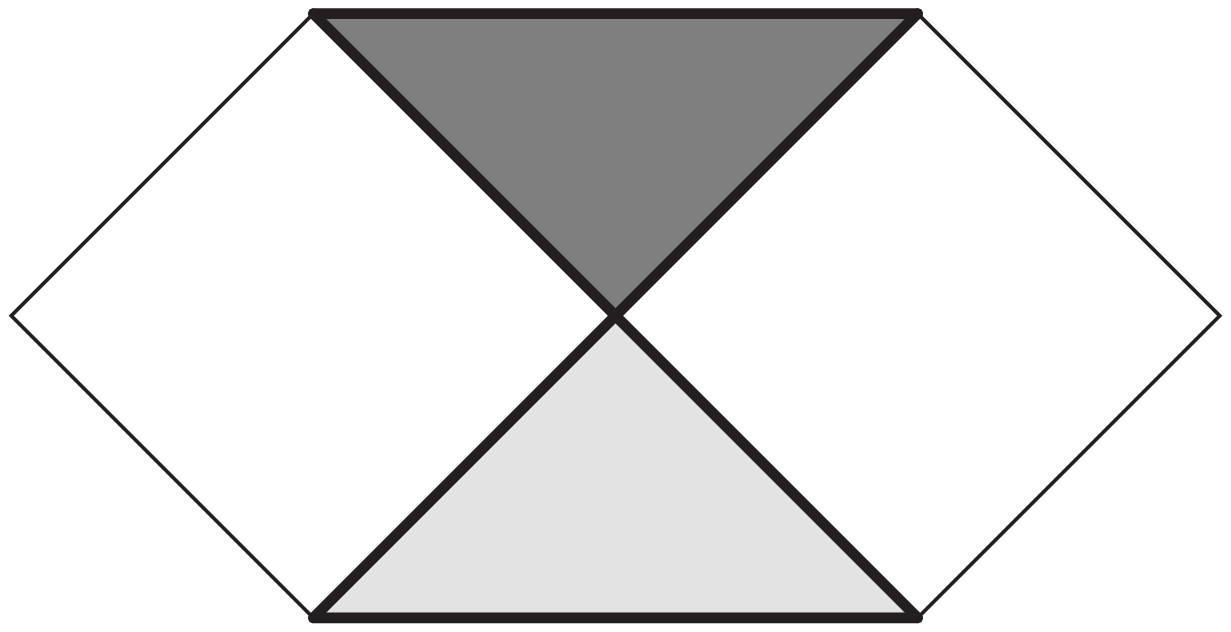}
\includegraphics[height=28mm]{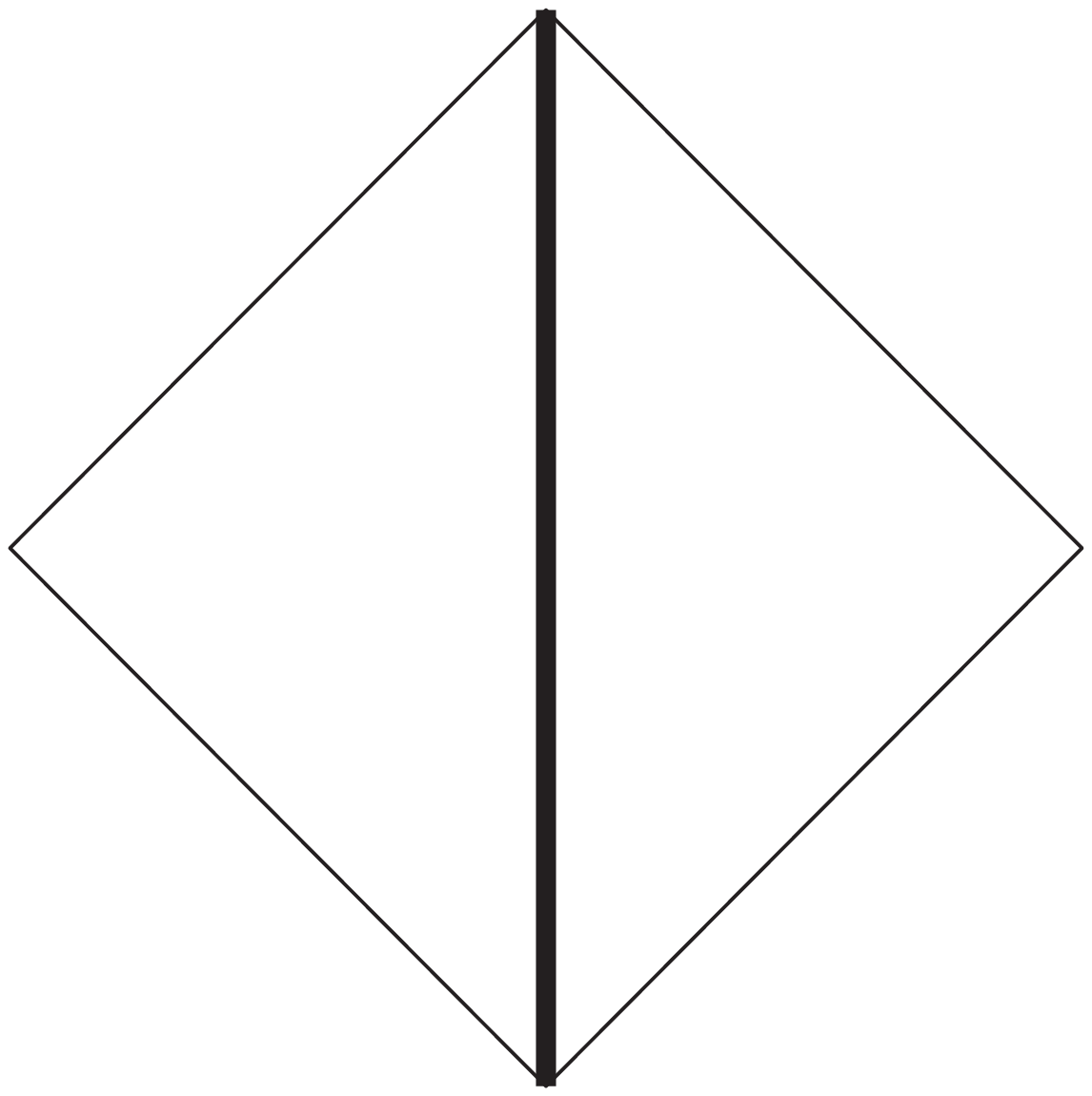}
 \vspace{-2mm}
 \caption{Penrose diagrams of (i) a CGHS black hole and (ii) an HKL traversable
 wormhole. The bold lines represent the trapping horizons, constituting the
 event horizons of the black hole and the throat of the wormhole.}
 \label{fig:bh-wh}
\end{figure}

In vacuum, $f=g=0$, the general solution to the field equations is \cite{CGHS}
\begin{equation}\label{vacuumsol}
  r = 2m - 4\lambda^2  x^+ x^-
\end{equation}
where the origin has been fixed. The constant $m$ may be interpreted as the
mass of the space-time, whose global structure has been described previously
\cite{cc}. For $m>0$ this describes the CGHS static black hole, analogous to
the Schwarzschild black hole. The Penrose diagram is shown in
Fig.~\ref{fig:bh-wh}(i).

Recently the solution \cite{dw}
\begin{equation}\label{staticwhsol}
  r = a + 2\lambda^2 (x^+ - x^- )^2,\qquad
  g = 2\lambda (x^+ - x^- ), \qquad
  f = 0
\end{equation}
has been found, where the origin has again been fixed. If $a>0$, this
represents a traversable wormhole, hereafter called the HKL wormhole, with
analogous global structure to a Morris-Thorne wormhole: a throat $r=a$ at
$x^+=x^-$, joining two regions with $r>a$, an $x^+>x^-$ universe and a
reflected $x^+<x^-$ universe, as depicted in Fig.~\ref{fig:bh-wh}(ii).

Thus the model naturally contains both static black holes and static
traversable wormholes. A characteristic feature of both cases is the trapping
horizons, defined by $\nabla r\cdot\nabla r=0$, or equivalently $\partial_+r=0$
or $\partial_-r=0$ \cite{wh,IH,1st,bhd}. In the CGHS black hole, they coincide
with the event horizons $r=2m$ at $x^-=0$ and $x^+=0$ respectively. In the HKL
wormhole, there is a double trapping horizon, $\partial_+r=\partial_-r=0$, at
the throat $r=a$. This illustrates how trapping horizons of different type may
be used to locally define both black holes and wormholes \cite{wh}. Also
relevant are the locally trapped regions where $\nabla r\cdot\nabla r<0$,
consisting of future trapped regions if $\partial_\pm r<0$ or past trapped
regions if $\partial_\pm r>0$, as occur in black holes or white holes
respectively. Locating the trapping horizons and the locally trapped regions is
a key feature of the analysis of dynamic situations.

In the following, we wish to take delta-function profiles for the radiation
energy densities
\begin{equation}
\rho_\pm=(\partial_\pm f)^2-(\partial_\pm g)^2
\end{equation}
where units have been fixed. To avoid ill-defined square roots of delta
functions, the model may be formally generalized to
\begin{eqnarray}
 \label{constraints1}
 \partial_\pm \partial_\pm r &=& -\rho_\pm\\
 \label{requation}
 \partial_+ \partial_- r &=& -4 \lambda^2\\
 \label{rhoequation}
 \partial_\pm \rho_\mp &=& 0
\end{eqnarray}
where the energy densities $\rho_\pm$ are now regarded as basic and need not be
derived from Klein-Gordon fields. The evolution equations
(\ref{requation}-\ref{rhoequation}) have the general solutions
\begin{eqnarray}
  \label{rsolution}
  r(x^+,x^-) &=& r_+ (x^+) + r_- (x^-) - 4\lambda ^2 x^+ x^-\\
  \label{rhosolution}
  \rho_\pm(x^+,x^-) &=& \rho_\pm (x^\pm).
\end{eqnarray}
The constraints (\ref{constraints1}) are preserved by the evolution equations
in the $\partial_\mp$ directions, and so may be reduced to
\begin{equation}\label{constraints}
  \partial_\pm \partial_\pm r_\pm = -\rho_\pm.
\end{equation}
The constraints are manifestly integrable for $r_\pm$ given initial data
\begin{eqnarray}\label{data}
 \rho_\pm\qquad&&\hbox{on $x^\mp=x_0^\mp$}\\
 (r,\partial_+r,\partial_-r)\qquad&&\hbox{at $x^+=x_0^+$,
$x^-=x_0^-$}
\end{eqnarray}
for constants $x_0^\pm$. The data consist of the energy-density profiles of the
left-moving and right-moving radiation, plus lower-dimensional data for the
metric. Then the general procedure is to specify this initial data according to
the desired physical situation, integrate the constraints (\ref{constraints})
for $r_\pm$, then the solution follows as (\ref{rsolution}-\ref{rhosolution}).
Consequently, the effect of the radiation is much easier to see than in
Einstein gravity, though the model shares various physically important features
including gravitational collapse to black holes satisfying cosmic censorship
\cite{cc}.

%%%%%%%%%%%%%%%%%%%%%%%%%%%%%%%%%%%%%%%%%%%%%%%%%%%%%%%%%%%%%%%%%%%%%%%%%%%%%%%%%%%%%%%
\section{construction of a wormhole from a black hole}
\label{sec:const} We study how to  construct a traversable wormhole from a
black hole by irradiating it with negative energy. Although a similar idea has
been studied previously \cite{dw}, now we present a simpler solution involving
impulsive radiation, which is a preliminary to construct an analytic solution
in four-dimensional Einstein gravity. We consider a CGHS black hole subjected
to impulsive negative-energy radiation at some positive value $x_0$ of the
Kruskal-like coordinates $x^{\pm}$, with energy density chosen in order to
close up the future trapped region by merging its trapping horizons, followed
by the constant irradiation needed to maintain the static wormhole:
\begin{eqnarray}
\rho_{\pm}&=&-4\lambda^2x_0\delta(x^{\pm}-x_0)-4\lambda ^2\Theta(x^{\pm}-x_0)
\end{eqnarray}
where $\Theta$ is the unit Heaviside step function and $\delta$ the Dirac
(delta-function) distribution. When differentiating to check solutions, it may
be useful to remember that the derivative $\delta'$ of the delta function, as a
distribution acting on test functions $f$, satisfies $\delta'f=-\delta f$ or
$(\delta f)'=0$. Of course $\Theta'=\delta$.

\begin{figure}
\includegraphics[height=5cm]{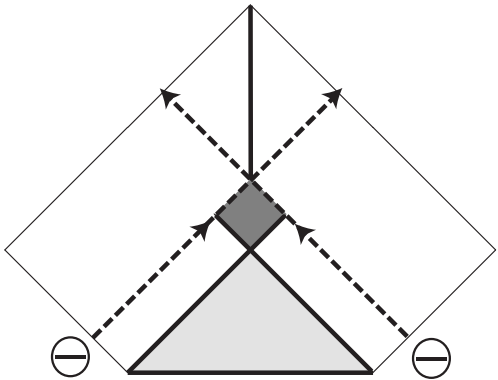}
 \vspace{-2mm}
 \caption{An HKL wormhole is constructed from a CGHS black hole by irradiating
 with impulsive negative-energy radiation, represented by dashed lines.  The
 bold lines represent the trapping horizons, light shading indicates past
 trapped regions and darker shading indicates future trapped regions. The
 impulses shift the black-hole trapping horizons suddenly, making them coincide,
 then constant non-impulsive radiation supports the resulting wormhole.}
 \label{fig:wh-creation}
\end{figure}

%The constraints become
%\begin{equation}
%\partial_{\pm}\partial_{\pm}r_{\pm}=
%4\lambda ^2\Theta(x^{\pm}-x_0)+4\lambda^2x_0\delta(x^{\pm}-x_0).
%\end{equation}
%Integrating twice,
Assuming a black hole of mass $m$ in the initial region, we obtain the solution
%\begin{equation}
%r_{\pm}=m+2\lambda
%^2(x^\pm-x_0)^2\Theta(x^{\pm}-x_0)+4\lambda^2x_0(x^{\pm}-x_0)\Theta(x^{\pm}-x_0)
%\end{equation}and
\begin{eqnarray}
r&=&2m -4\lambda^2x^+x^-+2\lambda^2({x^+}^2-{x_0}^2)\Theta(x^+-x_0)\nonumber\\
&&+2\lambda^2({x^-}^2-{x_0}^2)\Theta(x^--x_0).
\end{eqnarray}
The solution in the final region $x^\pm>x_0$ can be recognized as an HKL
wormhole (\ref{staticwhsol}). In more detail, the trapping horizons
\begin{equation}
0=\partial_{\pm}r=4\lambda^2\left(x^\pm\Theta(x^\pm-x_0)-x^\mp\right)
\end{equation}
are located at
\begin{equation}
\left\{
\begin{array}{ll}
x^\mp=0, &\qquad x^{\pm}<x_0\\
x^+=x^-,&\qquad x^{\pm}>x_0.
\end{array}
\right.
\end{equation}
and their radii are
\begin{equation}
r_0=\left\{
\begin{array}{ll}
2m, &\qquad x^{\pm}<x_0\\
2m-4\lambda^2x_0^2,&\qquad x^{\pm}>x_0.
\end{array}
\right.
\end{equation}
as depicted in Fig.~\ref{fig:wh-creation}. In this solution, the throat radius
$a=2m-4\lambda^2x_0^2$ of the wormhole is smaller than the horizon radius $2m$
of the initial black hole. Thus we require $2\lambda^2x_0^2<m$ in order that a
wormhole is constructed.

The results are similar to the previous solution \cite{dw}, except that here
the trapping horizons move discontinuously rather than continuously, a
well-known property under infinitesimally thin mass shells \cite{HE}. This is a
general feature of the solutions presented in this article, stemming from the
fact that the field equations or Einstein equations relate
$\partial_\pm\partial_\pm r$ to the energy densities $\rho_\pm$, so that
delta-function $\rho_\pm$ causes discontinuous $\partial_\pm r$. This is, of
course, an idealization of a situation where a concentrated packet of radiation
causes swift movement of the trapping horizon.

A recently discovered four-dimensional wormhole solution \cite{pr} can be
similarly constructed from a Schwarzschild black hole in full Einstein gravity
\cite{KHK}. By the time reverse, we can also obtain a picture where a wormhole
collapses into a black hole by beaming in impulsive radiation at the moment of
switching off the supporting ghost radiation. In this case, the horizon radius
of the black hole is larger than the throat radius of the initial wormhole.
This reduces to the sudden collapse case \cite{dw} without the impulsive
radiation and $x_0=0$.

%%%%%%%%%%%%%%%%%%%%%%%%%%%%%%%%%%%%%%%%%%%%%%%%%%%%%%%%%%%%%%%%%%%%%%%%%%%%%%%%%%%%%%%%%%%
\section{wormhole enlargement and reduction}\label{sec:ope}

\begin{figure}
\includegraphics[width=4cm]{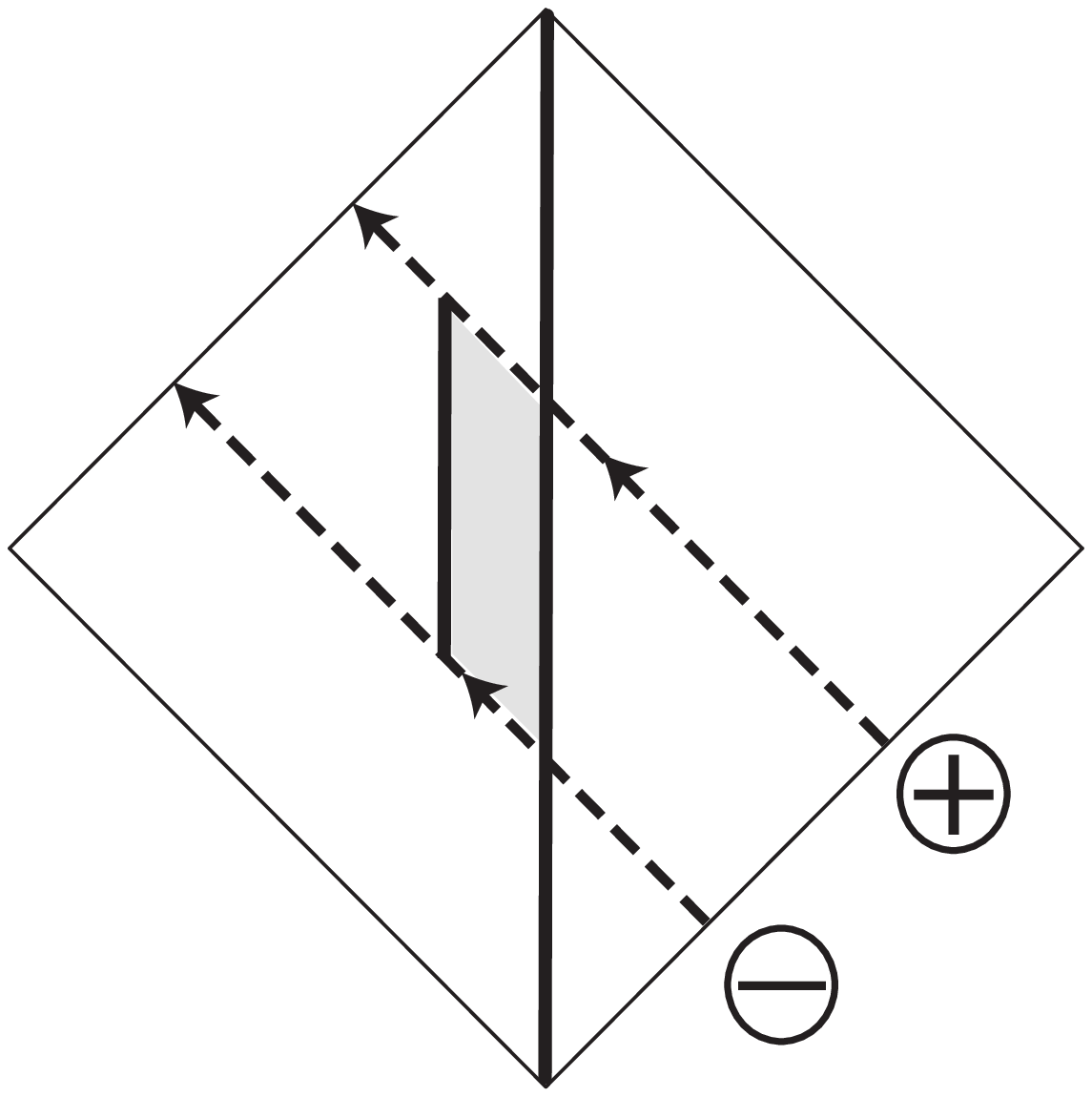}
\includegraphics[width=4cm]{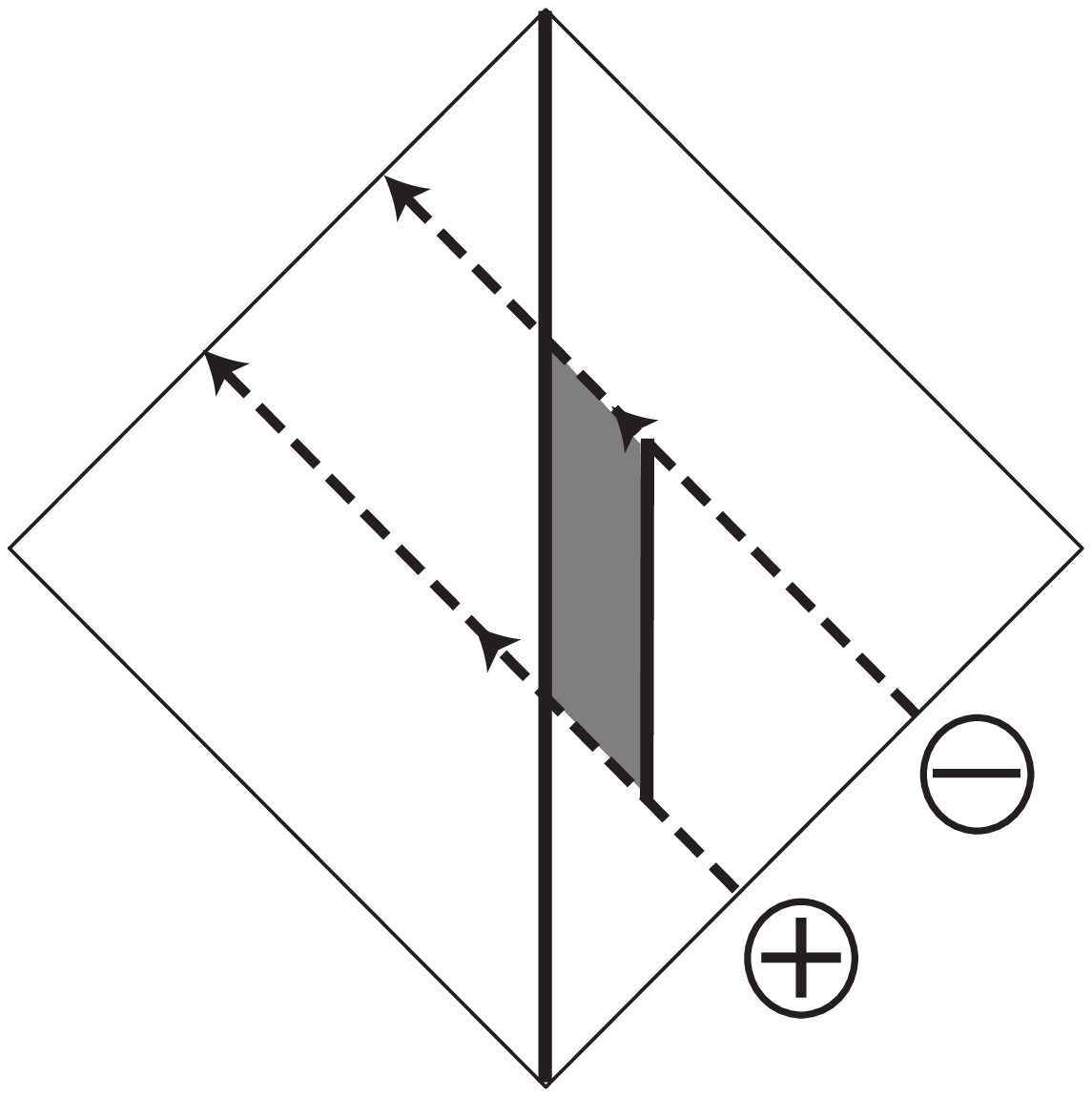}
 \vspace{-2mm}
 \caption{an HKL wormhole is subjected to a double burst of impulsive
 radiation with equal positive and negative energy. The operation shifts one
 wormhole mouth away from, then back to, its original position. The shaded
 regions are (i) past or (ii) future trapped, respectively expanding or contracting,
 so the wormhole becomes respectively larger or smaller.}
 \label{fig:whope}
\end{figure}

We are interested in how to create wormholes with throat large
enough for human beings to pass from one universe to another. It is practically
useful if it can be achieved by processes from our universe only. In this
section, we study wormhole operation by energy balance from one universe only.
We irradiate the wormhole from our universe with impulsive radiation of equal
positive and negative energy at different times:
\begin{eqnarray}
\rho_{+}&=&-4\lambda^2-\beta^2\delta(x^{+}-x_0)+\beta^2\delta(x^{+}-x_1)
\nonumber\\
\rho_{-}&=&-4\lambda^2
\end{eqnarray}
where $x_1>x_0$, so that the positive-energy impulse follows the
negative-energy impulse.
%The constraints
%\begin{eqnarray}
%\partial_{+}\partial_{+} r_{+}&=&4\lambda^2 -\beta^2\delta(x^{+}-x_1)+\beta^2\delta(x^{+}-x_0)\nonumber\\
%\partial_{-}\partial_{-} r_{-}&=&4\lambda^2
%\end{eqnarray}
%integrate to
%%%%%%%%%%%%%%%%%%%%%%%%
%\begin{eqnarray}
%r_{+}&=&2\lambda^2(x^+)^2+\frac{a}{2}-\beta^2(x^{+}-x_1)\Theta(x^{+}-x_1)+\beta^2(x^{+}-x_0)\Theta(x^{+}-x_0)\nonumber\\
%r_{-}&=&2\lambda^2(x^{-})^2+\frac{a}{2}
%\end{eqnarray}
Assuming a wormhole of throat radius $a$ initially, we obtain the solution
\begin{eqnarray}
r&=&a+2\lambda^2(x^+-x^-)^2+\beta^2(x^+-x_0)\Theta(x^+-x_0)\nonumber\\
&&-\beta^2(x^+-x_1)\Theta(x^+-x_1).
\end{eqnarray}
The locations of the trapping horizons $\partial_{\pm} r=0$ are given by
\begin{eqnarray}
x^-&=&\left\{\begin{array}{ll}
x^+,& \quad x^{+}<x_0\\
x^++\beta^2/4\lambda^2,&\quad x_0<x^+<x_1\\
x^+,& \quad x^{+}>x_1
\end{array}\right.\\
x^-&=&x^+.\nonumber
\end{eqnarray}
One sees that the wormhole mouth $\partial _+r=0$ is shifted out by the
negative-energy radiation and back again by the positive-energy radiation,
merging with the unmoved $\partial_-r=0$ mouth to leave a static wormhole
again, Fig.~\ref{fig:whope}(i). As in the previous section, the sudden shift of
the wormhole mouth is due to the impulsive nature of the radiation. The throat
radii of the initial and final wormholes are
\begin{equation}
r_0=\left\{\begin{array}{ll}
a,&\qquad x^{\pm}<x_0\\
a+\beta^2(x_1-x_0),&\qquad x^{\pm}>x_1
\end{array}
\right.
\end{equation}
so that the throat of the final state becomes larger than that of the initial
state.

On the other hand, if $x_1<x_0$, where the negative-energy impulse follows the
positive-energy impulse, the throat of the final state becomes smaller than
that of the initial state, Fig.~\ref{fig:whope}(ii). Then we need
$\beta^2(x_0-x_1)<a$ to obtain a wormhole rather than a naked singularity. If
one thinks of the positive-energy impulse as an idealized model of a
fast-moving traveller traversing the wormhole, this suggests how to operate and
maintain the wormhole for transport, including the back-reaction of the
traveller \cite{wh}. This also demonstrates the stability of the wormhole to
such dynamic perturbations.

In summary, the throat radius of the wormhole can be adjusted at will by
controlling the energy and timing of impulses. The example demonstrates the
property, following from the second law of wormhole dynamics \cite{wh}, that
the wormhole becomes smaller (respectively larger) after an operation in which
the wormhole mouths bifurcate to open up a contracting (respectively expanding)
region of future (respectively past) trapped surfaces and subsequently merge
again. Note that the energies $\pm\beta^2$ of the impulsive radiation are equal
and opposite, to ensure that the wormhole mouths merge again, as expected from
the first law of wormhole dynamics \cite{wh}.

%%%%%%%%%%%%%%%%%%%%%%%%%%%%%%%%%%%%%%%%%%%%%%%%%%%%%%%%%%%%%%%%%%%%%%%%%%%%%%%%%%
%%%%%%%%%%%%%%%%%%%%%%%%%%%%%%%%%%%%%%%%%%%%%%%%%%%%%%%%%%%%%%%%%%%%%%%%%%%%%%%%%%

\section{symmetric wormhole enlargement}\label{sec:enlarge}

\begin{figure}
\includegraphics[width=4cm]{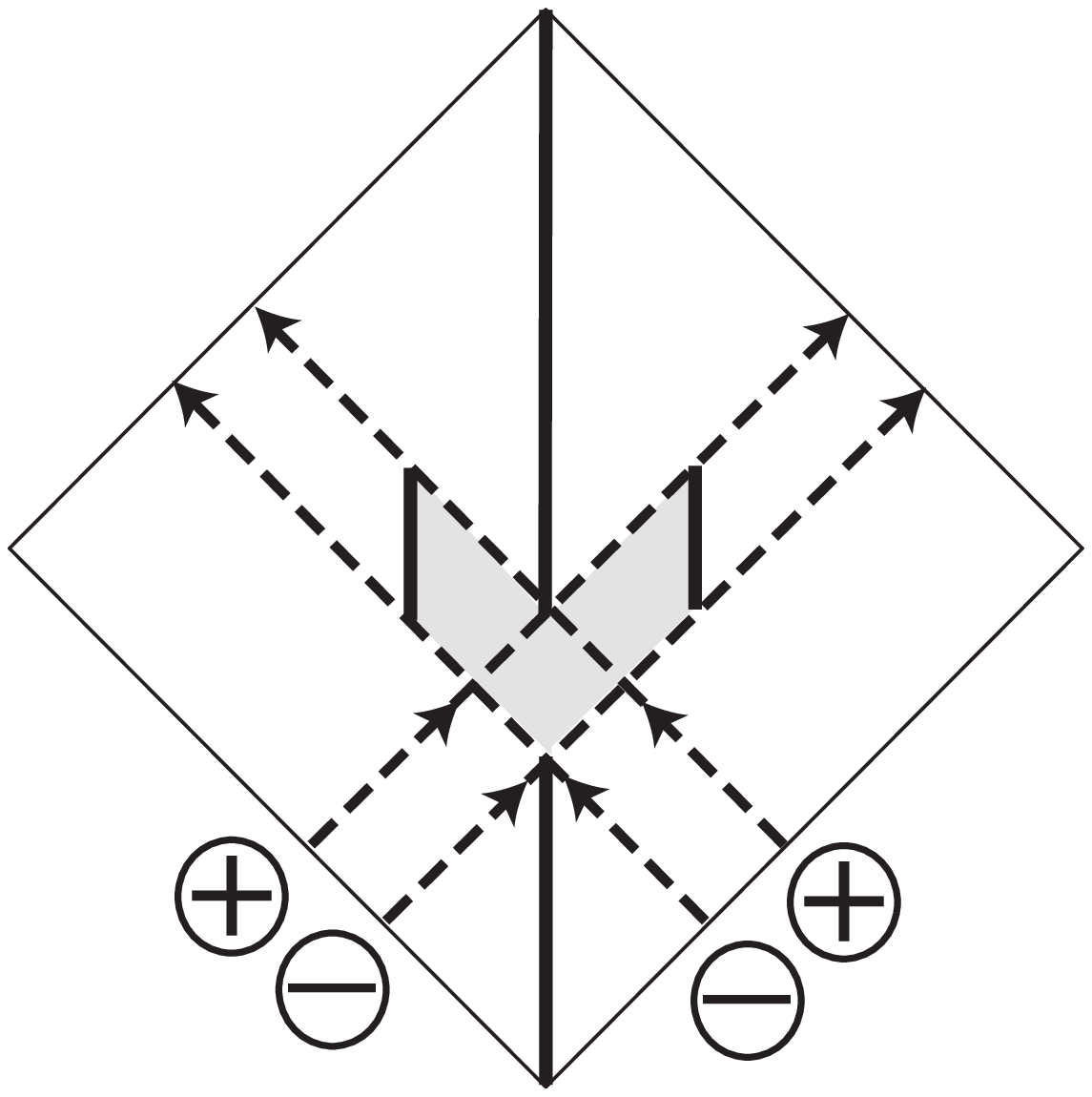}
\includegraphics[width=4cm]{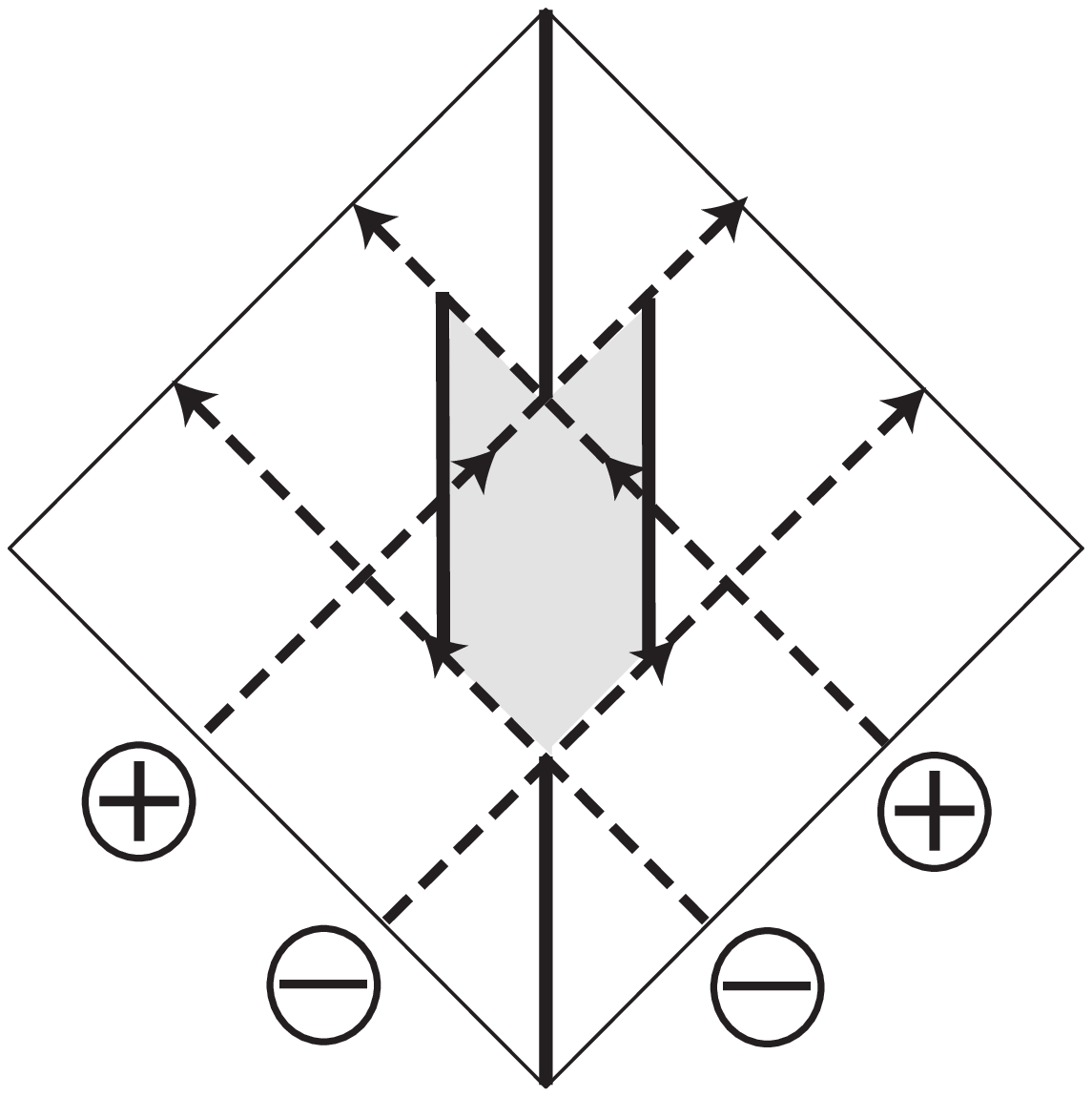}
 \vspace{-2mm}
 \caption{wormhole enlargement by symmetric bursts of impulsive radiation,
 keeping the non-impulsive radiation constant. Both wormhole mouths are
 shifted out, then back again, opening and closing an expanding (past trapped)
 region, shaded for a rapid burst (i) and a slow burst (ii).}
\label{fig:wh-enlarge1}
\end{figure}

Now we construct solutions to enlarge the throat by beaming in impulsive
radiation symmetrically from both universes. We give two solutions, a simple
one which is difficult to generalize to full Einstein gravity, and a more
delicate one which can be so generalized. The first example is a symmetrized
version of that of the previous section:
\begin{eqnarray}
\rho_{\pm}&=&-4\lambda^2-\beta^2\delta(x^{\pm}-x_0)+\beta^2\delta(x^{\pm}-x_1).
\end{eqnarray}
%Constraints are
%\begin{eqnarray}
%\partial_{\pm}\partial_{\pm}r_{\pm}=4\lambda^2+\beta^2\delta(x^{\pm}-x_0)
%-\beta^2\delta(x^{\pm}-x_1)
%\end{eqnarray}
%and integrating twice,
Assuming a wormhole with initial throat radius $a$, we obtain the solution
%\begin{eqnarray}
%r_{\pm}&=&2\lambda^2(x^{\pm})^2+\frac{a}{2}
%+\beta^2(x^{\pm}-x_0)\Theta(x^{\pm}-x_0)
%-\beta^2(x^{\pm}-x_1)\Theta(x^{\pm}-x_1)
%\end{eqnarray}
%and the solution is
\begin{eqnarray}
r&=&a+2\lambda^2(x^+-x^-)^2
+\beta^2(x^{+}-x_0)\Theta(x^{+}-x_0)\nonumber\\&&
+\beta^2(x^{-}-x_0)\Theta(x^{-}-x_0)
-\beta^2(x^{+}-x_1)\Theta(x^{+}-x_1)\nonumber\\&&
-\beta^2(x^{-}-x_1)\Theta(x^{-}-x_1).
\end{eqnarray}
This also describes an HKL wormhole (\ref{staticwhsol}) in the final region
$x^\pm>x_1$. The initial and final wormhole regions have throats
$\partial_+r=\partial_-r=0$ at $x^+=x^-$, with radii
\begin{eqnarray}
r_0&=&\left\{\begin{array}{ll}
a,&\qquad x^{\pm}<x_0\\
a+2\beta^2(x_1-x_0),&\qquad x^{\pm}>x_1\end{array} \right.
\end{eqnarray}
respectively. Then the wormhole is enlarged if $x_1>x_0$, so that the
negative-energy impulse precedes the positive-energy impulse, as before. The
final formation of a static wormhole is again dependent on the additional
radiation having equal and opposite energies $\pm\beta^2$. For a slow burst of
duration $x_1-x_0>\beta^2/4\lambda^2$, the wormhole mouths $\partial_\pm r=0$
are shifted so that the oppositely moving positive-energy impulses intersect
them, while for a rapid burst of duration $x_1-x_0<\beta^2/4\lambda^2$, the
mouths enclose the entire middle region $x_0<x^\pm<x_1$ between the impulses,
Fig.~\ref{fig:wh-enlarge1}.
% (Fig. \ref{fig:wh-enlarge1}).
%It is noting there is a past trapped surface in intermediate region
%between the initial and final wormhole regions (Fig. \ref{fig:wh-enlarge1}).

\begin{figure}
\includegraphics[width=7cm,height=7cm]{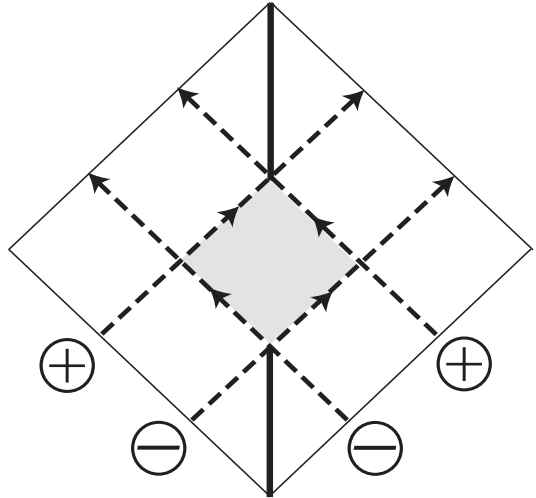}
 \vspace{-2mm}
 \caption{wormhole enlargement process by symmetric bursts of
 impulsive radiation, with negative energy followed by positive energy,
 timed as described in the text. The non-impulsive radiation is switched off
 between the impulses. Then the middle shaded region is vacuum and expanding.}
 \label{fig:wh-enlarge}
\end{figure}

%%%%%%%%%%%%%%%%%%%%%%%%%%%%%%%%%%%%%%%%%%%%%%
In full Einstein gravity, we wish to patch together Schwarzschild,
static-wormhole \cite{pr} and Vaidya regions. However, the four-dimensional
analogue of the middle region in the above example is not known analytically.
So for the second example, we switch off the non-impulsive radiation between
the impulses and switch it back on afterwards:
\begin{eqnarray}
\rho_{\pm}&=&
-4\lambda^2\Theta(x_0-x^{\pm})-\beta^2\delta(x^{\pm}-x_0)+\alpha^2\delta(x^{\pm}-x_1)
\nonumber\\&& -4\lambda^2\Theta(x^{\pm}-x_1).
\end{eqnarray}
Then the middle region is vacuum and therefore part of a CGHS white hole.
Assuming the initial region $x^{\pm}<x_0$ to be an HKL wormhole of throat
radius $a$, we find the solution
%constraints are
%\begin{eqnarray}
%\partial_{\pm}\partial_{\pm}r_{\pm}=4\lambda^2\Theta(-x^{\pm}+x_0)+\beta^2\delta(x^{\pm}-x_0)
%+4\lambda^2\Theta(x^{\pm}-x_1)-\alpha^2\delta(x^{\pm}-x_1).
%\end{eqnarray}
%Integrating twice,
\begin{widetext}
\begin{eqnarray}\label{gradualsol}
  r&=&a-4\lambda^2\left(x^+x^-+{x_0}^2\right)+4\lambda^2x_0(x^++x^-)
  +2\lambda^2(x^+-x_0)^2\Theta(x_0-x^+)+2\lambda^2(x^--x_0)^2\Theta(x_0-x^-)\nonumber\\
  &&+\beta^2(x^+-x_0)\Theta(x^+-x_0)+\beta^2(x^--x_0)\Theta(x^--x_0)
  -\alpha^2(x^+-x_1)\Theta(x^+-x_1)\nonumber\\&&-\alpha^2(x^--x_1)\Theta(x^--x_1)
  +2\lambda^2(x^+-x_1)^2\Theta(x^+-x_1)+2\lambda^2(x^--x_1)^2\Theta(x^--x_1).
 \end{eqnarray}
\end{widetext}
Now we want the final region $x^{\pm}>x_1$ to be an HKL wormhole in the usual
coordinates, Fig.~\ref{fig:wh-enlarge}.
%\begin{eqnarray}
%r_{\pm}&=&2\lambda^2(-x^{\pm}+x_0)^2\Theta(-x^{\pm}+x_0)
%+\beta^2(x^{\pm}-x_0)\Theta(x^{\pm}-x_0)-\alpha^2(x^{\pm}-x_1)\Theta(x^{\pm}-x_1)\nonumber\\
%&&+4\lambda^2x_0x^{\pm}+\frac{a}{2}-2\lambda^2x_0^2
%+2\lambda^2(x^{\pm}-x_1)^2\Theta(x^{\pm}-x_1).
%\end{eqnarray}
Then we find the relations
\begin{equation}
\label{relation} x_0=-\frac{\beta^2}{4\lambda^2},\qquad
x_1=-\frac{\alpha^2}{4\lambda^2}
\end{equation}
between the energy and timing of the impulses. This simplifies the solution in
the initial, middle and final regions:
\begin{eqnarray}\label{result}
r&=&\left\{\begin{array}{ll}
a+2\lambda^2(x^+-x^-)^2,&\quad x^{\pm}<x_0\\
a+4\lambda^2{x_0}^2-4\lambda^2x^+x^-,&\quad x_0<x^{\pm}<x_1\\
a+4\lambda^2(x_0^2-x_1^2)&\\\qquad+2\lambda^2(x^+-x^-)^2,&\quad
x_1<x^{\pm},\end{array} \right.
\end{eqnarray}
which are recognizable as CGHS (\ref{vacuumsol}) white-hole or HKL wormhole
(\ref{staticwhsol}) regions. The wormhole throats are at $x^+=x^-$ with radii
\begin{eqnarray}
r_0&=&\left\{\begin{array}{ll}
a,& \qquad x^{\pm}<x_0\\
a+4\lambda^2(x_0^2-x_1^2) ,& \qquad x_1<x^{\pm}.
\end{array}
\right.
\end{eqnarray}
%The diamond region $x_0<x^{\pm}<x_1$ is vacuum then a black hole of which the
%horizon radius is $a+\frac{\beta^4}{4\lambda^2}$.
%If the inequality
%\begin{eqnarray}\label{ineq}
%x_1>x_0+\frac{\beta^2}{2\lambda^2}
%\end{eqnarray}
%is satisfied, all of the vacuum region $x_0<x^{\pm}<x_1$ is covered by the past trapped surface, and
Thus the throat radius of the final wormhole is larger than that of the initial
one, again due to the negative-positive energy ordering of the burst.

Note that $x_0<x_1<0$, so that $\alpha^2<\beta^2$, meaning that the
positive-energy impulsive radiation does not completely balance the
negative-energy impulsive radiation, since some supporting negative energy has
already been removed. The difference $\alpha^2-\beta^2=-4\lambda^2(x_1-x_0)$
equals the energy $\int_{x_0}^{x_1}\rho_\pm dx^\pm$ missing as compared with
the static wormhole.

Combining with the results from the previous section, we have confirmed that
the radius of the wormhole throat is enlarged when an expanding region of past
trapped surfaces is opened and closed between the initial and final static
wormholes, by bifurcating and merging the wormhole mouths, defined as trapping
horizons. In addition, the solution just presented is one of the simplest where
the wormhole is enlarged, in the sense that each relevant region is part of
either a static white hole, a static wormhole or a pure-radiation region,
joined at null boundaries, which is also possible in full Einstein gravity
\cite{KHK}.

\section{summary}
\label{sec:summary} In this paper, we have used an exactly soluble dilaton
gravity model to study wormhole dynamics under impulsive radiation, finding
solutions where a traversable wormhole is created from a black hole or the
throat radius of a wormhole is enlarged or reduced, the size being controlled
by the energy and timing of the impulses. Where the solutions consist of
black-hole, static-wormhole and pure-radiation regions matched along null
boundaries, we can succeed to construct similar analytic solutions in full
Einstein gravity, though the analytical details are much more complex
\cite{KHK}.

The recipe to enlarge the wormhole is to cause the wormhole mouths to
bifurcate, opening up an expanding region of past trapped surfaces, then merge
again, by adding additional negative energy followed by compensating positive
energy. The general proof involves the first and second laws of wormhole
dynamics and is a future important work in the unified framework for black-hole
and wormhole dynamics \cite{wh}. The second law determines whether the area
increases or decreases, and the first law quantifies it in terms of the energy
supplied and work done.

The results in this paper show how to create a traversable wormhole of human
size in principle, if negative-energy matter can be controlled. Self-inflating
wormholes were discovered recently \cite{SH}, but the present solutions are the
first to describe stable wormhole enlargement. Clarifying the dynamical
behavior of wormholes is a quite attractive subject, since the cosmic
short-cuts and time travel usually considered as science fiction are thereby
closer to being realized.

\acknowledgements HK was supported in part by grant R01-2000-00015 from KOSEF,
SAH by Korea Research Foundation grant KRF-2001-015-DP0095, and SWK by Korea
Research Foundation grant KRF-2000-041-D00128.


\begin{references}
\bibitem{MT}M S Morris and K S Thorne, {Am. J. Phys.} {\bf 56}, 395 (1988).
\bibitem{V}M Visser, Lorentzian Wormholes: from Einstein to Hawking (AIP Press 1995).
\bibitem{HV1}D Hochberg and M Visser, {Phys. Rev.} {\bf D58}, 044021 (1998).
\bibitem{HV2}D Hochberg and M Visser, {Phys. Rev. Lett.} {\bf 81}, 746 (1998).
\bibitem{wh} S A Hayward, {Int. J. Mod. Phys.} {\bf D8}, 373 (1999).
\bibitem{IH}D Ida and  S A Hayward, {Phys. Lett.} {\bf A260}, 175 (1999).
\bibitem{bhd}S A Hayward, {Phys. Rev.} {\bf D49}, 6467 (1994).
\bibitem{1st}S A Hayward, {Class. Quantum Grav.} {\bf 15}, 3147 (1998).
\bibitem{CGHS} C G Callan, S B Giddings, J A Harvey and A Strominger,
 {Phys. Rev. D} {\bf 45}, R1005 (1992).
\bibitem{dw}S A Hayward, S-W Kim and H Lee, {Phys. Rev.} {\bf D65}, 064003 (2002).
\bibitem{SH}H A Shinkai and S A Hayward, {Phys. Rev.} {\bf D66}, 044005 (2002).
\bibitem{cc}S A Hayward, {Class. Quantum Grav.} {\bf 10}, 985 (1993).
\bibitem{HE}S W Hawking \& G F R Ellis,
The large scale structure of space-time, Cambridge University Press (1973).
\bibitem{pr}S A Hayward, {Phys. Rev.} {\bf D65}, 124016 (2002).
\bibitem{KHK}H Koyama, S A Hayward and S-W Kim, in preparation.
\end{references}
\end{document}